# VOCAL IDENTITY UNDER SIEGE BY AI VOICE CLONING TECHNOLOGIES

Jyh-An Lee[*] & Xuan Sun[**]

The advent of sophisticated AI-driven voice cloning has brought to the fore critical legal and ethical challenges regarding the protection of vocal identity. Prompted by recent controversies – including the striking resemblance between OpenAI's ChatGPT-4o voice and that of Scarlett Johansson – this article examines how generative AI technologies undermine the unique value of the human voice and further complicate the legal questions surrounding personal identity. Through a comparative analysis, the paper evaluates three principal legal frameworks: the right of publicity, personality rights, and the personal data protection right. Each framework – rooted in different legal traditions – offers distinct strengths and limitations in addressing the threats posed by AI-generated voice cloning. By analysing these doctrines' scope, remedies, and posthumous protections, the study offers a foundation for understanding how existing legal approaches may be applied to the evolving challenges of vocal identity in the era of generative AI.

## I. Introduction

In May 2024, OpenAI released an update to its artificial intelligence ("AI") chatbot, ChatGPT-4o, which featured a female voice called *Sky* interacting with users.[1] Many observers remarked upon the striking similarity of this voice to that of Scarlett Johansson in Spike Jonze's Oscar-winning 2013 dystopian science fiction film *Her*,

---

[*] Professor and Executive Director, Centre for Legal Innovation and Digital Society (CLINDS), The Chinese University of Hong Kong Faculty of Law.

[**] PhD Candidate, The Chinese University of Hong Kong Faculty of Law. Parts of earlier versions of this article were presented at the Intellectual Property and Technology in the 21st Century Conference at the Faculty of Law, National University of Singapore, the Intellectual Property and Human Creativity in the AI Age Conference at the UC Berkeley School of Law, the IP and Technology Law Seminar at Peking University Law School, the International Conference on Interdisciplinary Convergence and Collaborative Development of New Liberal Arts in the Age of Digital Intelligence at University of Science & Technology Beijing, as well as at the 17th Intellectual Property Conference – AI Beyond Imagination at the Faculty of Law, Chinese University of Hong Kong. The authors are grateful for the insightful and constructive comments offered by Songyin Bo, Albert Wai-Kit Chan, Chien-Yi Chang, Yang Chen, Yangzi Li, Jingwen Liu, Yinliang Liu, Rob Merges, Jingzhou Sun, David Tan, Qian Yin, Peter Yu, Peng Zhou, Anqi Wang, Runhua Wang, and participants at these conferences. This research was generously supported by a grant from the Research Grants Council of Hong Kong (Project No: CUHK 14608723).

[1] Andrew R Chow, "The Scarlett Johansson Dispute Erodes Public Trust In OpenAI", *Time* <https://time.com/6980710/scarlett-johansson-open-ai-sam-altman-trust/> (21 May 2024); Antonio Pequeño IV, "Sam Altman Apologizes to Scarlett Johansson Over OpenAI Chatbot Voice She Called 'Eerily Similar' To Hers", *Forbes* <https://www.forbes.com/sites/antoniopequenoiv/2024/05/21/sam-altman-apologizes-to-scarlett-johansson-over-openai-chatbot-voice-she-called-eerily-similar-to-hers/> (21 May 2024) [Pequeño IV].

in which Johansson voices a chatbot that becomes romantically involved with the protagonist.[2]

Johansson later issued a statement revealing that Sam Altman, OpenAI's chief executive, had approached her in September 2023, inviting her to provide the voice for *Sky* as a tribute to her role in the film – an offer which she declined.[3] She expressed that she was "shocked, angered and in disbelief" by subsequent developments.[4] The company asserted that the voice had not been modelled on Johansson's, nor had they sought to emulate her distinctive performance in *Her*.[5] Instead, the voice was generated by AI using samples from a different actor. Nonetheless, during the public demonstration of the *Sky* personal assistant, Altman posted a single-word tweet: "Her." Johansson interpreted this succinct message as an implicit acknowledgement, suggesting that the resemblance between Sky's voice and her own performance in *Her* was indeed intentional.[6] OpenAI eventually elected to remove the voice from service.[7] As a leading global AI company, questions have been raised as to whether OpenAI might have used Johansson's voice to train its AI model, resulting in a chatbot voice that bears a striking resemblance to hers.[8]

Although the dispute between Johansson and OpenAI did not result in litigation, it has nevertheless underscored the considerable value and identity interest inherent in the human voice. It further demonstrates how this value and identity may be undermined by the advent of AI technology, which is capable of readily cloning human voices. Moreover, this case is but one among many recent controversies involving AI-driven voice cloning. Individuals have used online voice generators to create fabricated audio clips purporting to show Emma Watson reading Adolf Hitler's *Mein Kampf*, Joe Biden announcing that US troops will enter Ukraine to combat Russia's invasion, and Star Wars actors uttering offensive, misogynistic, and racist remarks.[9]

---

[2] Bobby Allyn, "Scarlett Johansson Says She Is 'Shocked, Angered' over New ChatGPT Voice", *NPR* <https://www.npr.org/2024/05/20/1252495087/openai-pulls-ai-voice-that-was-compared-to-scarlett-johansson-in-the-movie-her> (20 May 2024) [Allyn]; Chow, *supra* note 1; Matt Murphy, "Scarlett Johansson 'Shocked' by AI Chatbot Imitation", *BBC News* <https://www.bbc.com/news/articles/cm559l5g529o> (21 May 2024) [Murphy]; Kat Tenbarge, "Scarlett Johansson says she was 'shocked, Angered' When She Heard OpenAI's Voice That Sounded Like Her", *NBC News* <https://www.nbcnews.com/tech/tech-news/scarlett-johansson-shocked-angered-openai-voice-rcna153180> (21 May 2024) [Tenbarge].

[3] Murphy, *supra* note 2; Pequeño IV, *supra* note 1; Sian Cain, "Scarlett Johansson Says OpenAI's Sam Altman Would Make a Good Marvel Villain After Voice Dispute", *The Guardian* <https://www.theguardian.com/film/article/2024/jul/18/scarlett-johansson-chatgpt-voice-openai-sam-altman> (17 July 2024); Tenbarge, *supra* note 2.

[4] Allyn, *supra* note 2; Chow, *supra* note 1; Murphy, *supra* note 2; Pequeño IV, *supra* note 1; Tenbarge, *supra* note 2.

[5] Allyn, *supra* note 2.

[6] "Scarlett Johansson Issues Statement About Rejecting Sam Altman's Request for Voice Work", *ChatGPT Is Eating the World* <https://chatgptiseatingtheworld.com/2024/05/20/scarlett-johansson-issues-statement-about-rejecting-sam-altmans-request-for-voice-work/> (20 May 2024).

[7] Chow, *supra* note 1; Murphy, *supra* note 2.

[8] Nicola Jones, "Who Owns Your Voice? Scarlett Johansson OpenAI Complaint Raises Questions", *Nature* <https://www.nature.com/articles/d41586-024-01578-4> (29 May 2024); Tenbarge, *supra* note 2.

[9] Tom Acres, "Extra Safeguards' Coming After AI Generator Used to Make Celebrity Voices Read Offensive Messages", *Sky News* <https://news.sky.com/story/extra-safeguards-coming-after-ai-generator-used-to-make-celebrity-voices-read-offensive-messages-12799703> (31 January 2023).

Such incidents are becoming increasingly prevalent. For instance, Gayle King, anchor of CBS's morning show, discovered that her voice had been misappropriated for use in a weight-loss advertisement,[10] while Greg Marston, a British voice actor with over two decades of experience, discovered his voice had been cloned by AI in an online demonstration.[11] These developments have given rise to a growing array of legal and policy challenges relating to vocal identity in the era of generative AI technologies.

This article explores the complex challenges that AI technologies pose to vocal identity and assesses how existing legal doctrines may be applied to address these challenges from a comparative law perspective. Following the introduction, Part II considers the human voice as a unique form of biometric information, analogous to facial features. It examines not only the vital role that voice plays in identity within social contexts, but also its capacity to reveal nuanced personal attributes through scientific inquiry. Part III discusses the extent to which AI technologies have blurred the boundaries of vocal identity, with increasingly prevalent AI-driven voice cloning undermining the distinctive role of the human voice. This phenomenon has given rise to a host of social, legal, and ethical dilemmas, affecting a broad spectrum of individuals, not solely celebrities. Part IV introduces the three legal rights relevant to the protection of vocal identity interests – namely, the right of publicity, personality rights, and the personal data protection right. Despite differing in origins and requirements, all three have been utilised to safeguard the integrity of vocal identity facing AI technologies. Part V undertakes a comparative analysis of these rights from a functional standpoint, identifying their shared objective of protecting identity interests. It also distinguishes between them in relation to subject, remedies, the necessity for actual use of personal data, and posthumous protection, considering how these distinctions may influence legal outcomes in the context of AI-generated voice cloning. Part VI concludes.

## II. Voice as Identity

Voice production is an innate and universal biological function in humans,[12] much like blinking or swallowing. Yet, the quality of an individual's voice possesses its own unique characteristics.[13] The distinctive configuration of each individual's vocal apparatus determines the qualities that set their voice apart from others. During the production of voiced sounds, the vibration of the vocal folds modulates the airflow passing through the glottis, thereby generating sound – the source of the voice. This sound travels through the vocal tract and is differentially amplified or attenuated at various frequencies.[14] Therefore, voice quality serves as a quasi-permanent characteristic for everyone.[15] Owing to this uniqueness, listeners can frequently identify a person solely by the quality of their voice.

---

[10] Nikolas Lanum, "Gayle King Fumes over Manipulated AI Video of Her Endorsing Weight Loss Company: 'Don't Be Fooled'", *Fox News* <https://www.foxnews.com/media/gayle-king-fumes-manipulated-ai-video-endorsing-weight-loss-company> (4 October 2023).

[11] *Ibid*.

[12] However, some people may lose the ability to produce a normal voice due to different reasons, such as diseases like Vocal Cord Paralysis. See *eg*, Oleksandr Butskiy, Bhavik Mistry & Neil K Chadha, "Surgical Interventions for Pediatric Unilateral Vocal Cord Paralysis: A Systematic Review" (2015) 141 JAMA Otolaryngology Head Neck Surgery 654 at 655.

[13] Robert J Podesva & Patrick Callier, "Voice Quality and Identity" (2015) 35 Ann Rev Applied Linguistics 173 at 173.

Unlike other forms of biometric information, such as fingerprint or iris patterns, the human voice possesses a distinctive capacity for recognition without the aid of technological assistance; we often identify people we know simply by hearing them speak. The voice carries a unique social recognisability, enabling others to associate a person with this attribute in everyday interactions. In this respect, the voice occupies a special category of biometric information – much like facial features – which typically involves personal attributes that are both physical and outwardly visible, and which defines an individual's appearance.[16] Such biometric attributes enable individuals to directly present and express their identities in social contexts, facilitating interpersonal recognition and connection in a way that is both immediate and personal.

The diverse qualities of voice perceived by human ears can be described with greater precision through a range of parameters, including frequency and intensity patterns, widths, shapes, slopes, mean frequencies, and the separations of the bars.[17] By comparing whether two voice samples share the same parameters, one can ascertain whether they originate from the same speaker, thus confirming identity. Today, this approach is widely applied in voice authentication systems[18] and forensic identification.[19]

The advent of digital technologies, including AI, has enabled us to discern a growing array of personal traits encoded within an individual's voice. While it is common to glean various types of personal information – such as a speaker's age, and gender – from the study of voice data,[20] advances in machine learning,[21] deep learning,[22] and other AI technologies now enable scientists to discern a speaker's emotional state through the analysis of their vocal features. By capturing the acoustic characteristics of speech, algorithms can effectively interpret the instinctive and intuitive emotions of the speaker.[23] For example, when someone is sad, there tends to be a slight elevation and wider variation in pitch compared to a neutral state.[24] Moreover, scientists have discovered that voice carries genetic information: as early as 2002, researchers identified the FOXP2 gene as linked to the human capacity for language, with particular mutations causing articulation difficulties.[25] In 2023, a variant in the ABCC9 gene was found to be correlated with a high-pitched voice. By analysing vocal acoustics, modern technology is now able to trace genetic origins and even infer other aspects of an individual's physical characteristics.

---

14 Zhaoyan Zhang, "Mechanics of Human Voice Production and Control" (2016) 140 J Acoustical Soc Am 2614 at 2614.

15 David Abercrombie, *Elements of General Phonetics* (Edinburgh: Edinburgh University Press, 2022) at 90.

16 Ellen S Bass, "A Right in Search of a Coherent Rationale-Conceptualizing Persona in a Comparative Context: The United States Right of Publicity and German Personality Rights" (2008) 42 USF L Rev 799 at 838; Joseph J Beard, "Clones, Bones and Twilight Zones: Protecting the Digital Persona of the Quick, the Dead and the Imaginary" (2001) 16 BTLJ 1165 [Beard] at 1265.

17 Bernard S Kamine, "The Voiceprint Technique: Its Structure and Reliability" (1969) 6 San Diego L Rev 213 at 216.

18 Hajer Y Khdier, Wesam M Jasim & Salah A Aliesawi, "Deep Learning Algorithms Based Voiceprint Recognition System in Noisy Environment" (2021) 1804 J Physics Conf Series 012042 at 1; Sarit K Mizrahi, "A Whole New Meaning to Having Our Head in the Clouds: Voice Recognition Technology, the Transmission of Our Oral Communications to the Cloud and the Ability of Canadian Law to Protect Us from the Dangers It Presents" (2017) 15 CJLT 121 at 122.

19 Harry Francis Hollien, *Forensic Voice Identification* (Cambridge: Academic Press, 2002).

20 Andreas Nautsch *et al*, "The GDPR & Speech Data: Reflections of Legal and Technology Communities: First Steps Towards a Common Understanding" in Proceedings of Interspeech 2019 (Graz, Austria: ISCA, 2019) 3695 at 3697 <https://doi.org/10.21437/Interspeech.2019-2647>.

21 *Ibid.*

22 Tae-Wan Kim & Keun-Chang Kwak, "Speech Emotion Recognition Using Deep Learning Transfer Models and Explainable Techniques" (2024) 14 Applied Sci 1553.

Finally, an individual's voice can reveal their state of health. For example, by extracting the paralinguistic features from patients with Parkinson's Disease (PD) and those without, and employing these as inputs for machine learning models, researchers have achieved an AUC score of 0.85 in cross-model performance for PD detection.[26] Google's research team has also introduced the HeAR AI model, designed to screen, diagnose, monitor and manage a wide variety of health conditions – including tuberculosis and chronic obstructive pulmonary disease – by capturing subtle clues hidden within bioacoustic signals.[27]

Given the identity-defining nature of human voice, it is possible for the voice qualities of two individuals with similar vocal apparatus configurations to sound alike. If misused, such similarity may give rise to legal issues concerning voice misappropriation. The interest in protecting identity via one's voice has been recognised by the law across numerous jurisdictions for decades.[28] For example, in *Midler v Ford Motor Co*, following a refusal by the celebrated singer Bette Midler, Ford Motor in the United States employed another singer whose voice was remarkably similar to Midler's, though far less renowned, to perform a song for their advertisement.[29] The US Court of Appeals for the Ninth Circuit remarked:

> A voice is as distinctive and personal as a face. The human voice is one of the most palpable ways identity is manifested. We are all aware that a friend is at once known by a few words on the phone. At a philosophical level it has been observed that with the sound of a voice, "the other stands before me." … A fortiori, these observations hold true of singing, especially singing by a singer of renown. The singer manifests herself in the song. To impersonate her voice is to pirate her identity.[30]

---

[23] K Tarunika, RB Pradeeba & P Aruna, *Applying Machine Learning Techniques for Speech Emotion Recognition*, 9th International Conference on Computing, Communication and Networking Technologies (ICCCNT), Bengaluru (2018) <https://ieeexplore.ieee.org/document/8494104/>.

[24] Serdar Yildirim *et al*, *An Acoustic Study of Emotions Expressed in Speech*, Proceedings of Interspeech, Jeju Island (2004) <https://www.isca-archive.org/interspeech_2004/yildirim04_interspeech.html>.

[25] Wolfgang Enard *et al*, "Molecular Evolution of FOXP2, a Gene Involved in Speech and Language" (2002) 418 Nature 869 at 869.

[26] John M Tracy *et al*, "Investigating Voice as a Biomarker: Deep Phenotyping Methods for Early Detection of Parkinson's Disease" (2020) 104 J Biomedical Informatics 103362 1 at 1, 9.

[27] Shravya Shetty, "This AI Model Is Helping Researchers Detect Disease Based on Coughs", *Google Blog* <https://blog.google/technology/health/ai-model-cough-disease-detection/> (19 August 2024).

[28] See *eg,* Seth E Bloom, "Preventing the Misappropriation of Identity: Beyond the Right of Publicity" (1990–91) 13 Hastings Comm & Ent LJ 489 at 499–505.

[29] *Midler v Ford Motor Co*, 849 F 2d 460 (9th Cir, 1988) [*Midler*].

[30] *Ibid* at 463.

While this is not the only case concerning the misappropriation of a celebrity's voice in the United States,[31] similar legal disputes have arisen in other jurisdictions, including France[32] and Germany.[33] Prior to the advent of AI technology, voice misappropriation typically occurred by way of human imitation, whereby one person would mimic another's voice, thereby creating a false impression of the voice owner's presence.[34] However, given the inherent uniqueness of each person's voice, such misappropriation by imitation was relatively uncommon. As a result, most celebrities could easily safeguard the identity interest vested in their voice.

## III. The Waning of Vocal Identity Amidst AI Technologies

However, AI technology has notably blurred the boundaries of vocal identity. Today, AI can easily clone the precise qualities of a speaker's voice by analysing and learning from the unique acoustic patterns of individuals.[35] AI voice cloning, also known as "verbal deepfakes", are generated using deep learning algorithms that produce synthetic speech, convincingly mimicking the nuances of authentic human voices.[36] AI models emulate distinct speech patterns and vocal cadences through systematic exposure to extensive recordings of human speech.[37] By analysing subtle nuances in tone, rhythm, and inflection, these models gradually learn to reproduce the unique characteristics that define an individual's manner of speaking with remarkable fidelity. This sophisticated process enables AI to generate synthetic voices that closely mirror the original, blurring the distinction between genuine and artificially created speech. In contrast to traditional imitation by another speaker with a similar

---

31 See *eg*, *Waits v Frito-Lay, Inc*., 978 F 2d 1093 (9th Cir, 1992) [*Waits*].

32 Logie, *infra* note 99.

33 OLG Hamburg, "Heinz Erhardt" (1989) GRUR 666 [*Heinz Erhardt*]; Huw Beverley-Smith, Ansgar Ohly & Agnès Lucas-Schloetter, *Privacy, Property and Personality: Civil Law Perspectives on Commercial Appropriation* (Cambridge: Cambridge University Press, 2005) at 126.

34 Elena Cooper, "AI and Performers' Rights in Historical Perspective" (2023) 45(8) EIPR 444 at 445; Leonard A Wohl, "The Right of Publicity and Vocal Larceny: Sounding Off on Sound-Alikes" (1988) 57 Fordham L Rev 445 at 446.

35 Please note that while there are legal issues beyond vocal identity concerning the unauthorised use of human voices for the purpose of AI training, these issues extend beyond the scope of this paper. For example, in 2023, the US Federal Trade Commission and the Department of Justice charged Amazon with enhancing its speech recognition system using voice recordings collected via its Alexa products ("FTC and DOJ Charge Amazon with Violating Children's Privacy Law by Keeping Kids' Alexa Voice Recordings Forever and Undermining Parents' Deletion Requests", *Federal Trade Commission* <https://www.ftc.gov/news-events/news/press-releases/2023/05/ftc-doj-charge-amazon-violating-childrens-privacy-law-keeping-kids-alexa-voice-recordings-forever> (31 May 2023)). Similarly, the European Union has raised concerns that designers of virtual voice assistants may seek to improve their products' performance by accessing stored voice snippets on these devices ("Guidelines 02/2021 on Virtual Voice Assistants" *European Data Protection Board* <https://www.edpb.europa.eu/system/files/2021-07/edpb_guidelines_202102_on_vva_v2.0_adopted_en.pdf> (2021) [European Data Protection Board]).

36 Edward Lee, "AI and the Sound of Music" (2024) 134 Yale LJF 187 [Lee] at 215; Elizabeth Shields, "The AI Doppelgänger Dilemma: Cloned Voices in the Music Industry" (2025) 48 Seattle UL Rev 761 [Shields] at 763.

37 Bryn Wells-Edwards, "What's in a Voice? The Legal Implications of Voice Cloning" (2022) 64 Ariz L Rev 1213 [Wells-Edwards] at 1214.

voice, such synthetic voices are significantly more convincing and potentially more manipulative.[38]

Early synthetic voice generation technology relied on converting text-strings into phonetics strings by pairing text input with pre-stored phonetic data.[39] For instance, the CHATR system – devised by the ATR Interpreting Telecommunications Research Laboratories in Kyoto, Japan – necessitates approximately one hour of recorded speech from the individual whose voice is to be emulated.[40] This recorded material is meticulously analysed and deconstructed into its most fundamental acoustic components, enabling the synthesis of new words and sentences that convincingly mimic the original speaker's voice.[41] As more voice samples were collected and larger databases established, this approach proved challenging to scale and was costly.[42]

With the continuous evolution of AI technology in synthetic voice generation, algorithms such as Hidden Markov Models ("HMMs")[43] and WaveNet[44] can now learn and replicate the speaking patterns of individuals. This advancement addresses previous limitations related to database size, enabling the generation of more natural and nearly indistinguishable synthetic voices at significantly lower costs. For instance, HMM-based speech synthesis allows context-dependent models to be trained from databases of natural speech, generating speech waveforms directly from the models.[45] This system facilitates the modelling of various speech styles with a much smaller recording database,[46] thus reducing the difficulty in producing highly misleading synthetic speech.

Meanwhile, the fully probabilistic and autoregressive WaveNet technology generates predictive distributions for each audio sample based on all preceding recordings.[47] The WaveNet model can accurately capture the characteristics of numerous speakers and switch between different identities,[48] making synthetic conversations more convincing and potentially deceptive. Today, users need only upload a few minutes of audio samples of their chosen speaker, upon which the AI analyses and constructs a sophisticated template of the studied voice.[49] The website Revoicer exemplifies an AI tool, capable of converting text into speech in forty languages and offering a wide array of intonations, moods, and styles.[50] Consequently, it has become remarkably straightforward for individuals to train their own AI models to replicate any voice, in virtually any language of their choosing.[51] This digital model can then be provided with textual scripts, skilfully manipulated, and seamlessly integrated into a wide range of platforms and interfaces.[52]

---

38 *Ibid* at 1213.

39 Debra Yarrington *et al*, *A System for Creating Personalized Synthetic Voices*, Proceedings of the 7th International ACM SIGACCESS Conference on Computers and Accessibility, Baltimore (2005) 196 at 197 <https://dl.acm.org/doi/10.1145/1090785.1090827>.

40 Beard, *supra* note 16 at 1192.

41 *Ibid*, Shannon Flynn Smith, "If It Looks Like Tupac, Walks like Tupac, and Raps like Tupac, It's Probably Tupac: Virtual Cloning and Postmortem Right-of-Publicity Implications" (2013) 2013 Mich St L Rev 1719 at 1727.

42 Heiga Zen *et al*, "The HMM-based Speech Synthesis System (HTS) Version 2.0" in 6th ISCA Workshop on Speech Synthesis (2007) 294 at 294 <https://www.cs.cmu.edu/~awb/papers/ssw6/ssw6_294.pdf>.

43 K Tokuda *et al*, "Speech Parameter Generation Algorithms for HMM-Based Speech Synthesis" in 2000 IEEE International Conference on Acoustics, Speech, and Signal Processing. Proceedings (Cat No 00CH37100) (2000, vol 3) 1315 at 1315 <http://ieeexplore.ieee.org/document/861820/>.

44 Aaron van den Oord *et al*, "WaveNet: A Generative Model for Raw Audio" (2016) <http://arxiv.org/abs/1609.03499>.

45 Zen *et al*, *supra* note 42.

46 *Ibid*.

47 Oord *et al*, *supra* note 44 at [1].

48 *Ibid*.

AI technologies represented by HMMs and WaveNet have enabled the more efficient production of voice-based content, such as audiobooks. For example, in May 2025, First Lady Melania Trump released the audiobook version of her memoir, *Melania*, utilising an AI-generated replica of her own voice – marking the first occasion on which a US First Lady has employed AI to narrate a commercial publication.[53] Film companies distributing their movies internationally can now utilise AI technology to translate each actor's voice into other languages, ensuring that it remains the actor's own voice, rather than that of a dubbing artiste.[54] Additionally, AI voice-cloning software has been employed to subtly alter an actor's accent, tailoring it to suit the preferences of audiences in the target market.[55] AI voice-cloning technologies are increasingly prevalent across a variety of domains, including marketing campaigns, advertising, and smart speaker applications.[56]

However, the widespread adoption of AI voice cloning technology has greatly facilitated the creation of identity confusion, leading to a significant increase in legal cases concerning voice misappropriation.[57] Increasingly, the voices of celebrities have been exploited for commercial purposes, as well as in political propaganda. Such uses become even more deceptive when cloned voices are incorporated into deepfake videos. For instance, in 2023, the renowned film star Tom Hanks appeared – without his consent – in a video endorsing a dental plan. This video, created using deepfake technology, prompted Hanks to publicly disclaim any association with the content via an Instagram announcement.[58]

Even distinguished politicians have fallen victim to these practices. During the election campaign for the 47th President of the United States, an AI-generated voice reproducing Joe Biden's distinctive acoustic features was weaponised by political adversaries. Steven Kramer, a former Louisiana Democratic political consultant, distributed robocalls using this synthetic Biden voice, urging recipients not to vote in New Hampshire's presidential primary in an attempt to manipulate the electoral process.[59]

---

49 Justine Phillips *et al*, "The Darwinian Effect: The Weaponization of Artificial Intelligence by Cyber Criminals" (2025) 61 Cal WL Rev 43 at 55.

50 Madhumita Murgia, "How Actors Are Losing Their Voices to AI", *Financial Times* <https://www.ft.com/content/07d75801-04fd-495c-9a68-310926221554> (1 July 2023) [Murgia].

51 Yang Chen, "Is Chinese Law Well-Prepared for AI Songs?: A Note of Caution on the Over-Expansion of Personality Rights" (2024) 42 Cardozo Arts & Ent LJ 261 [Chen] at 265.

52 Wells-Edwards, *supra* note 37 at 1214.

53 Vismaya V, "Melania Trump Uses AI to Narrate Her New Memoir", Decrypt <https://decrypt.co/321752/melania-trump-uses-ai-to-narrate-her-new-memoir> (23 May 2025).

54 Josh Rottenberg, "De-aged stars, Cloned Voices: How AI is Changing Acting", LA Times <https://www.latimes.com/entertainment-arts/movies/story/2025-07-24/hollywood-tomorrow-acting-jobs-ai-mark-hamill> (24 July 2025).

55 *Ibid*.

56 Wells-Edwards, *supra* note 37 at 1214–1215.

57 Lee, *supra* note 36 at 215.

58 The Guardian, "Tom Hanks says AI version of him used in dental plan ad without his consent", *The Guardian* <https://www.theguardian.com/film/2023/oct/02/tom-hanks-dental-ad-ai-version-fake> (1 October 2023).

The victims of such voice misuse are no longer limited to public figures. Nowadays, scammers can easily clone a voice using just three seconds of audio, misleading victims into believing that their loved ones or close friends are in urgent need of financial assistance.[60] In Hong Kong, a company suffered multimillion-dollar losses in a deepfake scam, where the Chief Financial Officer's image and voice were cloned to participate in a video conference call. Following the call, an employee transferred approximately US$25.6m to the scammer.[61] Similar incidents of financial fraud, facilitated by increasingly sophisticated verbal deepfake technology, have been reported in various countries across the globe.[62]

Moreover, in Maryland, United States, a high school principal's voice was imitated by an athletic director after the principal declined to renew his contract. The synthetic recording, which contained racist remarks, was uploaded to the internet, provoking widespread outrage against the principal.[63] His denial of any association with the recording was ultimately substantiated when FBI forensic analysts concluded that it contained traces of AI-generated content, with human editing to add realistic background noises.[64] The fact that synthetic racist recording was produced at minimal cost illustrates the growing legal risk posed by AI voice cloning technologies.[65] Meanwhile, the principal's need to rely on FBI expertise to prove the recording was not his own underscores the complexity and expense involved in determining whether a voice is authentically of human origin or artificially generated.

---

[59] Maggie Astor, "Political Consultant Who Orchestrated Fake Biden Robocalls Is Indicted", *The New York Times* <https://www.nytimes.com/2024/05/23/us/politics/biden-robocalls-steve-kramer-democratic-primary.html> (23 May 2024).

[60] Anna Desmarais, "AI scams can now impersonate your voice. Here's how to avoid them", *EuroNews* <https://www.euronews.com/next/2025/07/13/ai-scams-can-now-impersonate-your-voice-heres-how-to-avoid-them?> (13 July 2025).

[61] Heather Chen & Kathleen Magramo, "Finance Worker Pays Out $25 Million After Video Call with Deepfake 'Chief Financial Officer'", *CNN* <https://www.cnn.com/2024/02/04/asia/deepfake-cfo-scam-hong-kong-intl-hnk/index.html> (4 February 2024).

[62] Charles Bethea, "The Terrifying A.I. Scam that Uses Your Loved One's Voice", *New Yorker* <https://www.newyorker.com/science/annals-of-artificial-intelligence/the-terrifying-ai-scam-that-uses-your-loved-ones-voice> (7 March 2024); Emily Flitter & Stacy Cowley, "Voice Deepfakes Are Coming for Your Bank Balance", *The New York Times* <https://www.nytimes.com/2023/08/30/business/voice-deepfakes-bank-scams.html> (30 August 2023); Catherine Stupp, "Fraudsters Used AI to Mimic CEO's Voice in Unusual Cybercrime Case" *Wall Street Journal* <https://www.wsj.com/articles/fraudsters-use-ai-to-mimic-ceos-voice-in-unusual-cybercrime-case-11567157402> (30 August 2019).

[63] Emilia David, "Gym Teacher Accused of Using AI Voice Clone to Try to Get a High School Principal Fired", *The Verge* <https://www.theverge.com/2024/4/25/24140556/ai-voice-cloning-baltimore-school-misuse> (26 April 2024).

[64] Christian Olaniran & Jessica Albert, "Local News School Principal Was Framed Using AI-Generated Racist Rant, Police Say. A Co-Worker Is Now Charged.", *CBS News* <https://www.cbsnews.com/baltimore/news/maryland-framed-principal-racist-ai-generated-voice/> (26 April 2024).

[65] Resemble AI, "Pricing That Scales with You", *Resemble AI* <https://www.resemble.ai/pricing/>.

## IV. Legal Rights Protecting the Identity Interest in Human Voice

The identity interest in voice is protected across various jurisdictions through differing legal approaches. Although misappropriation of voice may, at times, give rise to liabilities under rights or interests other than those specifically related to voice,[66] this article focuses on those legal rights that directly safeguard the identity interest in voice – namely, the right of publicity in the United States, personality rights in civil law jurisdictions, and personal data protection rights developed across various jurisdictions. While these three rights may occasionally overlap,[67] they nonetheless represent distinct approaches to the protection of vocal identity.

The enforcement of the three rights has responded to the challenges posed by AI to the protection of vocal identity in the past few years. For example, a court in the State of New York, United States, has held that the unauthorised use of synthetic voices closely resembling those of the voiceover actors, constitutes a violation of the voice owner's right of publicity. Similarly, in China – a civil law jurisdiction – personality rights are employed to shield vocal identity from the unauthorised use of recognisable synthetic voices. The General Data Protection Regulation ("GDPR") in the European Union (EU) stands as the most prominent personal data protection regime and has been applied to voice data as biometric information.

### A. *Right of Publicity*

The right of publicity is the principal cause of action invoked for claims of unconsented voice use in the United States. Under US common law, the right of publicity serves to protect individuals against the unauthorised appropriation of their

[66] For instance, in the aforementioned 'Joe Biden robocall' case, Steve Kramer, who created and disseminated the synthetic voice of Joe Biden, was indicted in New Hampshire on charges of felony voter suppression and misdemeanour impersonation of a candidate ("Steven Kramer Charged with Voter Suppression Over AI-Generated President Biden Robocalls", *New Hampshire Department of Justice* <https://www.doj.nh.gov/news-and-media/steven-kramer-charged-voter-suppression-over-ai-generated-president-biden-robocalls> (23 May 2024)). In the 'Maryland framed principal' case, the former high school athletic director, who generated the synthetic racist recording, entered a plea deal and was sentenced to four months in prison after facing charges that included theft, stalking, disruption of school operations, and retaliation against a witness (Moodee Lockman & Dennis Valera, "Baltimore man accused of framing school principal with an AI-generated rant takes a plea deal" , *CBS News* <https://www.cbsnews.com/baltimore/news/maryland-pikesville-principal-ai-voice-impersonation-trial/> (28 April 2025); Christian Olaniran & Jessica Albert, "School principal was framed using AI-generated racist rant, police say. A co-worker is now charged", *CBS News* <https://www.cbsnews.com/baltimore/news/maryland-framed-principal-racist-ai-generated-voice/> (April 26 2024)). In both instances, the legal interests underlying the causes of action extended beyond the identity interest in voice. See also Rebecca J Roberts, "You're Only Mostly Dead: Protecting Your Digital Ghost from Unauthorized Resurrection" (2023) 75 Fed Comm LJ 273 at 286–287 (analysing criminal liabilities arising from unauthorised voice cloning); Wells-Edwards, *supra* note 37 at 1224–1236 (discussing causes of action such as defamation and privacy tort of false light).

[67] For example, in European civil law countries, vocal identity is safeguarded both as a personality right under civil law and as a data right under the provisions of the GDPR. Similarly, in China, vocal identity is protected as a personality right in the Civil Code and as personal information under the Personal Information Protection Law.

identity – which includes name, likeness and voice – for the benefit of others.[68] The right of publicity, broadly defined as the "inherent right of every human being to control the commercial use of his or her identity",[69] has been well-established in the US for over sixty years.[70] However, it tends to be confined to the protection of identities with a commercial value, with the result that ordinary individuals who are not celebrities are unable to avail themselves of this right.[71] Its recognition of a proprietary interest in the identity of a well-known individual is analogous to the interest in goodwill or reputation protected in common law claims of passing off.[72] Section 46 of the Restatement (Third) of Unfair Competition seeks to clarify this concept, providing that:

> One who appropriates the commercial value of a person's identity by using without consent the person's name, likeness, or other indicia of identity for purposes of trade is subject to liability for the relief appropriate under the rules stated in §§ 48 and 49.[73]

The right of publicity is protected under common law in some states, by statute in others, and sometimes both.[74] Currently, more than thirty states in the United States recognise the right of publicity.[75] For a claim to succeed, usually the plaintiff must establish the use in question was for a commercial purpose – a criterion explicitly set out in the publicity statutes or common law of several states. However, in the absence of a federal right of publicity, the precise provisions vary from state to state. In some states, misappropriation must be for advertising or merchandising purpose,[76] whilst others generally prohibit any unauthorised use of an individual's identity, including news, books or political campaigns.

Fundamentally, the right of publicity is regarded as a property right in personality,[77] as it is inextricably linked to an individual's personhood.[78] Infringements upon the right of publicity are deemed commercial torts, frequently described as "unfair

---

[68] Olivia Wall, "A Privacy Torts Solution to Postmortem Deepfakes" (2023) 100 Wash U L Rev 885 at 892.

[69] J Thomas McCarthy & Roger E Schechter, *The Rights of Publicity and Privacy* (2nd ed, 2000) (May 2025 update) § 3:1 [McCarthy & Schechter].

[70] It was first recognised by the Second Circuit in 1953 that baseball players had a "right of publicity" in their images. *Haelan Laboratories Inc v Topps Chewing Gum Inc*, 202 F 2d 866, 868 (2nd Cir, 1953).

[71] See, *eg, DeClemente v Columbia Pictures Industries Inc*, 860 F Supp 30 (Dist Ct, Eastern District of New York, 1994); *Pesina v Midway Manufacturing Co*, 948 F Supp 40 (Dist Ct, Northern District of Illinois, 1996).

[72] David Tan, "The Fame Monster Reloaded: The Contemporary Celebrity, Cultural Studies and Passing Off" (2010) 32 Sydney L Rev 335 at 336; David Tan, *The Commercial Appropriation of Fame: A Cultural Analysis of the Right of Publicity and Passing Off* (Cambridge: Cambridge University Press, 2017) [Tan, *Fame*] at 199–245.

[73] Restatement (Third) of Unfair Competition (US: American Law Institute, 1995), § 46.

[74] In 2025, 33 states had provided their citizens with a remedy for infringement of the right of publicity; some by statute, some by common law and in some states by both sources of law (McCarthy and Schechter, *supra* note 69 at § 6:2).

[75] Jennifer E Rothman, *The Right of Publicity Privacy Reimagined for a Public World* (Cambridge: Harvard University Press, 2018) [Rothman] at 3.

[76] McCarthy & Schechter, *supra* note 69 at § 3:2.

[77] Rothman, *supra* note 75 at 1.

[78] Margaret Jane Radin, "Property and Personhood" (1982) 34 Stan L Rev 957 at 966.

competition". Accordingly, this right embodies elements of both property law and tort law.[79]

With the rapid development of modern advertising and communications throughout the mid-twentieth century, the right of publicity emerged as a distinct legal concept, gradually diverging from the right to privacy.[80] Far from being merely a variant of trademark, copyright, false advertising, or privacy rights, the right of publicity constitutes a distinct legal category. Although it shares certain affinities with these neighbouring areas of law, it possesses unique legal dimensions and underlying rationales that set it apart.[81]

Most states have enshrined the right of publicity within their statutory framework, expressly including "voice" among the protected attributes.[82] This recognition underscores the growing appreciation of an individual's vocal identity as a distinctive and valuable element, affording individuals legal recourse against the unauthorised commercial exploitation of their unique voice. There have been several cases in which the voice has been expressively protected under the right of publicity. A particularly illustrative example is *Waits v Frito-Lay, Inc*, in which the renowned musician Tom Waits brought an action against Frito-Lay for both misappropriation and false endorsement, following the company's use of a Waits sound-alike in a Doritos commercial.[83] The US Court of Appeals for the Ninth Circuit determined that Waits' voice sufficiently identifies him. Accordingly, Waits was able to pursue a claim under the right of publicity for the unauthorised commercial imitation of his voice.[84] The court upheld the jury's finding that Frito-Lay had infringed upon Waits's right of publicity by broadcasting the commercial featuring the impersonated voice.[85]

In 2024, Tennessee amended its Personal Rights Protection Act of 1984, introducing the "Ensuring Likeness, Voice, and Image Security Act of 2024" (ELVIS Act).[86] The ELVIS Act serves as a notable example of state legislation which does not restrict the protection of the right of publicity solely to advertising or merchandising purposes.[87] Furthermore, the Act renders a person liable to civil action "if the person publishes, performs, distributes, transmits, or otherwise makes available to the public an individual's voice or likeness, with knowledge that use of the voice or likeness was not authorized by the individual" or "if the person distributes, transmits, or otherwise makes available an algorithm, software, tool, or other technology, service, or device, the primary purpose or function … [of which] is the production of a particular, identifiable individual's photograph, voice, or likeness" without authorisation from that individual.[88]

In essence, under the ELVIS Act, the use of AI technology to clone an individual's voice without permission may itself give rise to civil liability. The ELVIS

---

[79] McCarthy & Schechter, *supra* note 69 at § 3:1.

[80] Melville B Nimmer, "The Right of Publicity" (1954) 19 L Contemp Probs 203 at 203–204.

[81] McCarthy & Schechter, *supra* note 69 at § 3:1.

[82] Beard, *supra* note 16 at 1190-1191; Chen, *supra* note 51 at 266–267.

[83] *Waits, supra* note 31, abrogated by *Lexmark Int'l, Inc. v. Static Control Components, Inc.*, 134 S. Ct. 1377 (2014).

[84] *Ibid* at 1098; see also *Midler, supra* note 29.

[85] *Ibid* at 1098–1102.

[86] Ensuring Likeness, Voice, and Image Security Act of 2024 (Tennessee, US) [ELVIS Act], Tenn Code Ann, §§ 47-25-1101 to 47-25-1108.

[87] ELVIS Act, Tenn Code Ann, §§ 47-25-1105(a).

[88] *Ibid*.

Act represents a pioneering piece of legislation, the first of its kind in the nation, designed specifically to utilise the right of publicity as a means of safeguarding artists from the unauthorised use of AI-generated voice clones, and extending the scope of protection beyond celebrities and well-known individuals. Therefore, commentators regard the ELVIS Act as both a potential blueprint for future legislative initiatives in other states and a significant experimental model that may inform the development of federal regulations governing the dissemination of unauthorised AI voice clones.[89]

In 2025, the United States District Court for the Southern District of New York rendered its decision in the case of *Lehrman, et al v Lovo, Inc*.[90] This case was brought by two voice actors, who discovered that their voices had been appropriated without their consent to train an AI algorithm and produce synthetic audio products. Among other causes of action, the claimants based their arguments on §§ 50 and 51 of the New York Civil Rights Law,[91] which codifies the right of privacy in State of New York.[92] The case highlights the unique vulnerability of vocal performers in the age of AI and raises important questions surrounding the protection of vocal identity under the right of publicity in the digital era.

### B. *Personality Right*

The personality right is a fundamental private right in civil law jurisdictions, designed to protect not only the essential attributes of natural persons[93] but also their economic or patrimonial interests, as most notably exemplified by the commercial exploitation of celebrity image rights.[94] Deeply rooted in the principles of human dignity, autonomy, and freedom of will,[95] this right encompasses a broad spectrum of personal interests. Commonly protected components include life, bodily integrity, name, likeness, image, reputation, and privacy, amongst others.[96]

---

[89] Shields, *supra* note 36 at 781–782.

[90] *Lehrman et al v Lovo, Inc*, Case 1:24-cv-03770-SPO, (10 July 2025) (Dist Court, Southern District of New York) [*Lovo*].

[91] *Ibid* at 25–26.

[92] New York Civil Rights Law 1903, §§ 50–51 (McKinney 2025). New York technically does not recognise the right of publicity. The statutory right of privacy allows an individual to bring a claim for unauthorised uses for advertising or trade purposes, "the name, portrait, picture, likeness, or voice" of that individual.

[93] Adrian Popovici, "Personality Rights – A Civil Law Concept" (2004) 50 Loy L Rev 349 at 351–352.

[94] Lilian Edwards & Edina Harbinja, "Protecting Post-Mortem Privacy: Reconsidering the Privacy Interests of the Deceased in a Digital World" (2013) 32 Cardozo Arts & Ent LJ 83 [Edwards & Harbinja] at 101.

[95] Bass, *supra* note 16 at 828, 838; Yang Chen, "Navigating the Identity Thicket in China from a Comparative Lens: Conflicting Control Rights over a Person's Name" (2023) 53 Hong Kong LJ 843 at 848; Chen, *supra* note 51 at 267–268; Susanne Bergmann, "Publicity Rights in the United States and Germany: A Comparative Analysis" (1999) 19 Loy LA Ent LJ 479 [Bergmann] at 503; Edward J Eberle, "Human Dignity, Privacy, and Personality in German and American Constitutional Law" (1997) 1997 Utah L Rev 963 [Eberle] at 979; Ryan Kraski, "Combating Fake News In Social Media: U.S. and German Legal Approaches" (2017) 91 St John's L Rev 923 at 931–932; Hannes Rösler, "Dignitarian Posthumous Personality Rights – An Analysis of U.S. and German Constitutional and Tort Law" (2008) 26 BJIL 153 [Rösler] at 168–169; Giorgio Resta, "The New Frontiers of Personality Rights and the Problem of Commodification: European and Comparative Perspectives" (2011) 26 Tul Eur & Civ LF 33 [Resta] at 50; Paul M Schwartz & Karl-Nikolaus Peifer, "Prosser's Privacy and the German Right of Personality: Are Four Privacy Torts Better than One Unitary Concept?" (2010) 98 Cal L Rev 1925 [Schwartz & Peifer] at 1948.

Notably, the personality right has been occasionally invoked to safeguard the human voice even before the advent of AI technologies. For instance, in 1989, the Higher Regional Court of Hamburg ruled that, akin to the protection afforded to an individual's likeness and name, the personality right also extends to the human voice, shielding it from unauthorised exploitation by other parties.[97] The court emphasised that imitating a celebrity's voice constitutes an illicit exploration of their personality, a protection which endures even after death.[98]

There have been similar court decisions in France recognising vocal identity as part of personality rights. For example, in May 1989 the defendant Schol released "L'Echo Dechavanne" in northern France and Belgium, featuring new beat music and samples from Christophe Dechavanne's talk show, edited to imply self-insult and encouragement of drug use.[99] In 1990, the Tribunal de Grande Instance of Lille ruled that a person's voice is a personal attribute protected against unauthorised use likely to cause confusion or harm, establishing civil liability. This judgment was upheld by the Douai Court of Appeal in 1992.[100]

One illustrative example relating to AI voice cloning and personality is China's 2020 Civil Code, which explicitly recognises voice as a distinct element of individual identity protected under personality rights.[101] A landmark decision in 2024 by the Beijing Internet Court further elucidated this legal protection. The case concerned a professional voice actor whose voice had been used, without her consent, to train an AI model that subsequently generated a text-to-speech product featuring her unique vocal characteristics.[102] The court held that the defendants' unauthorised commercial use of the synthetic speech, which preserved the plaintiff's distinct acoustic features, constituted an infringement of her personality right.

In reaching its decision, the court meticulously compared the timbre, intonation, and pronunciation style between the plaintiff's prior audio works and the synthetic voice. It concluded that the synthetic voice was sufficiently distinctive to enable recognition of the plaintiff by a particular group of listeners,[103] thereby falling within the scope of legally protectable voice interests. Consequently, even though one of the defendants used only the synthetic voice, rather than recordings of the plaintiff's actual voice, this was still found to have infringed her personality rights in relation to her voice.

---

[96] Bass, *supra* note 16 at 829–838; Bergmann, *supra* note 95 at 504–512; Chen, *supra* note 51 at 267–268; Thibault Gisclard, "Limitations of Autonomy of the Will in Conventions of Exploitation of Personality Rights" (2014) 45(1) IIC 18 at 19; Anne Lauber-Rönsberg, "The Commercial Exploitation of Personality Features in Germany from the Personality Rights and Trademark Perspectives" (2017) 107 The Trademark Reporter 803 [Lauber-Rönsberg] at 807; Eric H Reiter, "Personality and Patrimony: Comparative Perspectives on the Right to One's Image" (2002) 76 Tul L Rev 673 [Reiter] at 680–681; Rösler, *supra* note 95 at 181; Resta, *supra* note 95 at 49–50; Schwartz & Peifer, *supra* note 95 at 1967.

[97] *Heinz Erhardt*, *supra* note 33.

[98] *Ibid*.

[99] Philippe Logie, "The Dechavanne Case: Unauthorised Sound Sampling of a Distinctive Voice" 1993 4(4) Ent LR 121 [Logie] at 123.

[100] *Ibid* at 121, 123.

[101] Civil Code of the People's Republic of China (adopted 28 May 2020, effective 1 January 2021) [Civil Code 2020], Art 1023(2): "For the protection of a natural person's voice, the relevant provisions on the protection of the right to likeness shall be applied *mutatis mutandis*"; Art 1018: "a likeness is an external image reflected on a certain carrier by means of video, sculpture, and painting, among others, through which a specific natural person can be identified."

[102] *Yin v Zhongguang Broadcasting (ZB) et al* [2023] Beijing Internet Court Civ 12142 [*Yin*].

[103] *Ibid.*

It is evident that personality rights will assume an increasingly vital role in safeguarding vocal identity against the challenges posed by AI-driven voice cloning within civil law jurisdictions. As advances in generative AI continue to blur the boundaries between authentic and synthetic voices, the nuanced protection afforded by personality rights offers an essential legal framework to address unauthorised exploitation and misappropriation of vocal attributes. The recognition of the human voice as a distinct and protectable aspect of individual identity not only reinforces the principles of dignity and autonomy but also enables individuals – particularly those whose livelihoods depend on the integrity of their vocal identity – to seek meaningful redress.

### C. *Personal Data Protection Right*

Voice and other forms of biometric information are now being increasingly recognised and afforded protection as personal data in a growing number of jurisdictions. In countries lacking explicit laws on the right of publicity or personality rights, data protection legislation often serves as the primary safeguard for vocal identity.[104] This heightened recognition stems from the unique and immutable nature of biometric identifiers; such information is inherently linked to an individual's physical and behavioural characteristics, making it both highly personal and readily traceable.[105] Unlike passwords or other conventional forms of identification, features such as one's face, voice, or fingerprints cannot be easily altered – except, perhaps, through extreme surgical intervention.[106] Moreover, biometric data allows for direct personal identification without the need for supplementary information, in contrast to other sensitive data where, for example, possession of a credit card number alone does not necessarily reveal identity or facilitate fraud.[107] Therefore, biometric data is often regarded as the most intrinsically "personal" of all categories of personal data, meriting enhanced legal safeguards and careful stewardship.[108]

With the growing prevalence and sophistication of biometric technologies, the protection of biometric data emerged as a topic of considerable importance during the 2000s.[109] In response to these developments, the European Commission undertook two public consultations between 2009 and 2011, exploring the future evolution of the data protection framework and, specifically, the integration of biometric data protection within it.[110] These efforts ultimately culminated in the adoption of the GDPR in 2016,[111] which offers a comprehensive definition of 'biometric data'. According to the GDPR, biometric data refers to "personal data resulting from specific technical processing relating to the physical, physiological or behavioural characteristics of a natural person, which allow or confirm the unique identification of that natural person, such as facial images or dactyloscopic (fingerprint) data."[112] Given that voice data may be deployed to uniquely identify an individual, it unequivocally falls within the ambit of biometric data as envisaged by the GDPR. Personal data protection laws across the world have increasingly recognised voice data as a significant and distinct category of protected personal information.[113]

---

[104] Murgia, *supra* note 50.

[105] Derek Tu, "Proving Biometric Data Privacy Harm in Federal Courts: Borrowing Informational Harm Concepts from Common Law and Trade Secret" (2024) 52 AIPLA QJ 731 at 737.

[106] *Ibid*, Tunca Bolca, "Can PIPEDA 'Face' the Challenge? An Analysis of the Adequacy of Canada's Private Sector Privacy Legislation Against Facial Recognition Technology" (2020) 18 CJLT 51 [Bolca] at 66.

[107] Bolca, *supra* note 106 at 166.

[108] *Ibid*.

[109] Catherine Jasserand, "Legal Nature of Biometric Data: From 'Generic' Personal Data to Sensitive Data: Which Changes Does the New Data Protection Framework Introduce?" (2016) 2 EDPL 297 at 298.

The processing of personal data is regulated by Article 5 of the GDPR.[114] To lawfully collect and process personal data, the activity must be founded upon one of the legitimate bases enumerated under Article 6(1) of the GDPR.[115] The GDPR distinguishes between two categories of biometric data: the first pertains to physical and physiological human characteristics, such as weight, dactyloscopic data, eye colour, voice and ear shape recognition; the second concerns behavioural characteristics, including keystroke analysis, handwritten signature analysis, and eye tracking. Both categories have the potential to identify individuals. When considering the lawful processing of biometric data – which falls under the special categories of personal data – there is an additional layer of complexity when such data is processed for the purpose of uniquely identifying a human individual. Article 9 of the GDPR stipulates that, in order to process such data within the bounds of the law, both the requirements set out in Article 6 and the further stringent conditions set forth in Article 9 must be satisfied. Specifically, Article 9 prohibits the processing of "biometric data for the purpose of uniquely identifying a natural person," as well as other special categories of personal data, save for narrowly defined exceptions.[116]

---

[110] *Ibid* at 298–299.

[111] "The History of the General Data Protection Regulation", *European Data Protection Supervisor* <https://www.edps.europa.eu/data-protection/data-protection/legislation/history-general-data-protection-regulation_en>.

[112] European Commission, Regulation 2016/679 of 27 April 2016 on the Protection of Natural Persons with Regard to the Processing of Personal Data and on the Free Movement of Such Data, and Repealing Directive 95/46/EC (General Data Protection Regulation) [2016] OJ, L 119/1 [GDPR], Art 4.

[113] Stacy-Ann Elvy, "Age-Appropriate Design Code Mandates" (2024) 45 U Pa J Intl L 953 at 966–967, 974; Eugenia Georgiades and James Birt, "Georgiades and Birt, "Unravelling the Metaverse Matrix: Navigating Privacy Protection Within Modelling and Simulation Platforms" (2025) 20 Washington Journal of Law, Technology & Arts 102, 128; Carla Llaneza, "An Analysis on Biometric Privacy Data Regulation: A Pivot Towards Legislation Which Supports the Individual Consumer's Privacy Rights in Spite of Corporate Protections" (2020) 32 St Thomas L Rev 177 [Llaneza] at 187–190; Liubov Kirzhakova, "Facial Recognition Technology in the Market: What Consumers Need to Know to Protect Their Rights" (2024) 16 UC Law Science & Technology Journal 1 at 17–18; Mackenzie K Mendolla, "A Blurry Lens: Assessing the Complicated Legal Landscape of Biometric Privacy Through the Perspective of Mobile Apps" (2024) 54 Seton Hall L Rev 923 at 947–948, 953; Anvitha Sai Yalavarthy, "Aadhaar: India's National Identification System and Consent-Based Privacy Rights" (2023) 56 Vand J Transnat'l L 619.

[114] GDPR, *supra* note 112, Art 5.

[115] *Ibid*, Art 6

[116] *Ibid*, Art 9.

The European Data Protection Board has expressly stated in its Guidelines 02/2021 on Virtual Voice Assistants that "voice data is inherently biometric personal data".[117] The protection of voice data has been actively reflected in the enforcement of the GDPR. For instance, in 2022, the Hungary Supervisory Authority imposed an administrative fine of €650,000 on a Hungarian bank for its unlawful processing of voice data, which was used to ascertain the emotion state of customers during telephone conversations.[118] In the same year, the Spanish Data Protection Authority levied a fine of €6,000 on a company for the unlawful collection and processing of images and voices captured by video cameras.[119]

Although, to date, there have been no GDPR cases addressing the protection of vocal identity specifically against AI-driven voice cloning, legal scholars have widely acknowledged the applicability of the Regulation.[120] Moreover, in 2021, the European Parliament issued a policy paper, Tackling Deepfakes in European Policy,[121] noting that the proliferation of deepfake technologies, such as voice cloning, has contributed to a rise in crimes involving identity theft. As the creation of deepfakes commonly entails the use of biometric data, including voice fragments, the GDPR is accordingly engaged to regulate deepfake software and applications,[122] voice cloning technology among them.

## V. Comparing Three Approaches to Vocal Identity

Although the aforementioned three rights are underpinned by distinct rationales and, in some instances, originate from differing legal traditions, each serves to protect the identity interests inherent in an individual's voice. Given the considerable challenges to vocal identity presented by contemporary AI technologies, it becomes especially pertinent to examine the convergences and divergences between these three approaches when confronted with such issues. Accordingly, the following section undertakes a comparative analysis of the strengths and limitations associated with each of these legal frameworks, with particular reference to their capacity to address and respond to the novel challenges engendered by advancements in AI.

---

117 European Data Protection Board, *supra* note 35 at [13].

118 "Data protection issues arising in connection with the use of Artificial Intelligence", *European Data Protection Board* <https://www.edpb.europa.eu/news/national-news/2022/data-protection-issues-arising-connection-use-artificial-intelligence_en> (8 February 2022).

119 "Spain: AEPD fines unnamed company €6,000 for lack of legal basis for processing personal data", *Data Guidance* <https://www.dataguidance.com/news/spain-aepd-fines-unnamed-company-6000-lack-legal-basis?> (19 October 2022).

120 Shields, *supra* note 36 at 782–783.

121 "Tackling Deepfakes in European Policy", *European Parliament* <https://www.europarl.europa.eu/RegData/etudes/STUD/2021/690039/EPRS_STU(2021)690039_EN.pdf> (July 2021).

122 *Ibid* at [38]–[39].

### A. *Identity Interests*

As all three rights safeguard the identity interests of individuals, they form the central focus of this article, which specifically examines the vocal identity as it currently faces unprecedented challenges posed by AI cloning technologies.

The right of publicity enables individuals to assert proprietary interests in their identity, most notably by providing the means to monetise such identity – a facet of particular significance for celebrities and public figures.[123] Since its inception in 1953, the scope of the right of publicity has progressively broadened. Originally confined to a few markers of identity, such as name and likeness, its reach has since extended to other attributes, including one's voice or even a distinct catchphrase.[124] This expansive interpretation is also supported by insights from cultural studies into the making of contemporary celebrity, with widespread recognition of the unique characteristics and semiotic significance attached to such identities.[125] In *Midler*, the US Court of Appeals for the Ninth Circuit noted that actions arising under the right of publicity are particularly pertinent to celebrities whose identities possess significant commercial value.[126]

Importantly, the US courts have consistently underscored that it is the protection of identity itself – rather than the discrete, person-specific components – that lies at the very heart of the right of publicity. For instance, in the aforementioned *Lovo* case, the court elucidated this point, noting:

> [T]he statute is designed to protect a person's identity, not merely a property interest in his or her "name," "portrait or "picture." What matters is not whether, for example, a face is seen in a particular picture, but rather "the quality and quantity of the identifiable characteristics displayed," and whether they suffice to permit recognition of the plaintiff's identity.[127]

With respect to the personality right, its conceptual foundation was first articulated by Karl Gareis in 1877,[128] and has since been significantly enriched and refined by subsequent jurists.[129] Otto von Gierke, for example, emphasised that a defining feature of the personality right – setting it apart from other rights – is its inherently personal nature.[130] As a result, protection of the individual's identity has emerged as a fundamental aspect of the personality right.[131] Today, an increasing number of civil law scholars define personality or persona as the "indicia of identity", generally encompassing a person's name, likeness, and voice, among other attributes.[132]

---

[123] Bass, *supra* note 16 at 806–813, 817–820; Daniel J Gervais & Martin L Holmes, "Fame, Property & Identity: The Purpose and Scope of the Right of Publicity" (2014) 25 Fordham IP Media & Ent LJ 181 at 183; Shields, *supra* note 36 at 768.

[124] Marshall Leaffer, "The Right of Publicity: A Comparative Perspective" (2007) 70 Alb L Rev 1357 at 1362.

[125] Tan, *Fame*, *supra* note 72 at 64.

[126] *Midler*, *supra* note 29 at 463.

[127] *Lovo*, *supra* note 90 at 48.

[128] Karl Gareis, "*Die Privatrechtssphären im modernen Kulturstaate, insbesondere im Deutschen Reiche*" [The Private Law Spheres in the Modern Cultural State, Especially in the German Reich] (1877) 3 *Zeitschrift für Gesetzgebung und Praxis auf dem Gebiete des Deutschen öffentlichen Rechtes* 137; Karl Gareis, "*Das juristische Wesen der Autorrechte, sowie des Firmen-und Markenschutzes*" [The Juridical Nature of Author's Rights as well as of Trade Name and Trade-Mark Protection] (1877) 35 *Archiv für Theorie und Praxis des Allgemeinen und Deutschen Handels-und Wechselrechts* 185.

[129] Johann Neethling, "Personality Rights: A Comparative Overview" (2005) 38 Comp & Int'l LJS Afr 210 at 210–211.

Specific facets of the personality right, such as the right to one's name or image, have even been acknowledged by the courts as the 'rights to identity'. For example, the Italian Supreme Court has expressively recognised the right to personal identity as an integral component of the personality right. The court held that an individual's ascertainable cultural, professional, religious, political, and social experiences must not be distorted, misrepresented, falsified, confused, or otherwise contested by the attribution of false' – albeit not necessarily defamatory – statements or acts.[133] A similar emphasis on the centrality of identity can be found in other civil law jurisdictions that have embraced the framework of personality rights. In the landmark decision of *Aubry v Éditions Vice-Versa Inc*, the Supreme Court of Canada underscored this point, holding that "[t]here is an infringement of a person's right to his or her image and, therefore, fault as soon as the image is published without consent and enables the person to be identified."[134] These judicial interpretations collectively affirm that the essence of the right of personality lies in the protection of an individual's identity interests, thereby underscoring its continued relevance amidst evolving legal and technological landscapes.

Personal data protection laws are, by their very nature, designed to safeguard an individual's identity interests regardless of whether there is any commercial harm to the individual. In academic discourse, personal data is often defined as personally identifiable information.[135] This conceptualisation is reflected in the personal data laws of many jurisdictions. For instance, the GDPR defines "personal data" as "any information relating to an identified or identifiable natural person ('data subject')".[136] Similarly, China's Personal Information Protection Law ("PIPL") defines "personal information" as "all kinds of information related to identified or identifiable natural persons that are electronically or otherwise recorded, excluding information that has been anonymized."[137] The Personal Data (Privacy) Ordinance in Hong Kong similarly defines "personal data" as any information relating directly or indirectly to a living individual whose identity can be directly or indirectly ascertained, and which is accessible or capable of being processed.[138] Case law has accordingly established that whether particular data constitutes "personal data" depends on whether the identity of an individual can be traced from that data.[139] In short, the protection of identity interests is inherent to the concept of personal data, and forms the very subject matter of the personal data protection right.

---

[130] Neethling, *supra* note 129 at 211.

[131] Chen, *supra* note 51 at 267–268; Karen Eltis, "Is 'Truthtelling' Decontextualized Online Still Reasonable? Restoring Context to Defamation Analysis in the Digital Age" 63 McGill LJ 553 at 567–568; Keenan C Fennimore, "Reconciling California's Pre, Post, and Per Mortem Rights of Publicity" (2012) 22 Ind Intl & Comp L Rev 377 [Fennimore] at 399–402; Resta, *supra* note 95 at [61]; Schwartz & Peifer, *supra* note 95 at 1967.

[132] Andrew McGee & Gary Scanlan, "Phantom Intellectual Property Rights" (2000) 3 IPQ 264 [McGee & Scanlan] at 265, 268.

[133] Giorgio Pino, "The Right to Personal Identity in Italian Private Law: Constitutional Interpretation and Judge-Made Rights" in Mark van Hoecke & François Ost, eds, *The Harmonization of Private Law in Europe* (United Kingdom: Hart Publishing, 2000) 225 at 226.

[134] *Aubry v Éditions Vice-Versa Inc* [1998] 1 SCR 591 at [53] (SC, Canada).

[135] Maria Lilla Montagnani & Mark Verstraete, "What Makes Data Personal" (2023) 56 UC Davis L Rev 1165 at 1187.

[136] GDPR, *supra* note 112, Art 4.

[137] Personal Information Protection Law of the People's Republic of China 2021 [Personal Information Protection Law 2021], Art 4.

Taken together, the right of publicity, the personality right, and the personal data protection right each place the protection of individual identity at their core, albeit through varying legal frameworks and doctrinal emphasis. As the boundaries of what constitutes protected identity continue to expand – encompassing not only traditional elements such as name and image, but also more nuanced markers like voice – these rights become ever more essential in an age where technological advancements, such as AI-driven vocal cloning, pose new and complex threats.

### B. *Subject of the Rights*

The subjects protected by these three rights differ notably: the right of publicity is most associated with celebrities, whereas both the personality right and the personal data protection right extend equally to non-celebrities. This distinction stems from the differing origins and underlying rationales of the three rights.

Infringement of the right of publicity is a commercial tort of unfair competition in the United States, evidenced by the right's inclusion in the Restatement (Third) of Unfair Competition.[140] The right of publicity is thus fundamentally designed to safeguard the commercial value of one's identity.[141] Consequently, it is typically the names or likenesses of celebrities – rather than those of non-celebrities – that possess sufficient commercial value to warrant protection under this right.

Although, in principle, the right of publicity in the United States extends to all individuals,[142] judicial attention has predominantly centred upon celebrities.[143] This focus arises from the greater impetus celebrities possess to pursue legal actions when their identities are commercially exploited, in contrast to non-celebrities.[144] Accordingly, in *Midler*, the court explained that a claimant may seek redress under the right of publicity statute if their voice is distinctive, widely recognised, and has been deliberately imitated for the purpose of promoting a product.[145] By contrast, the voice of a non-celebrity typically lacks these characteristics, and thus does not attract the same legal protection, even though the voice of that individual may be synthetically copied to perpetuate scams. Furthermore, the reality that only celebrities seek to vindicate the right of publicity bears upon certain states' prerequisites for this quintessential right, which require that a defendant's use of the plaintiff's name or likeness be for advertising or merchandising purpose.[146]

---

[138] Personal Data (Privacy) Ordinance (Cap 486) (HK), s 2.

[139] Jojo Mo, "Justification for a disclosure order: *Shi Tao v Privacy Commissioner for Personal Data*" (2009) 20(7) Ent LR 270 at 270–271.

[140] Restatement (Third) of Unfair Competition, *supra* note 73, §§ 46–49.

[141] *Ibid*, § 38: "[o]ne who causes harm to the commercial relations of another by appropriating the other's intangible trade values is subject to liability to the other for such harm only if . . . the actor is subject to liability for an appropriation of the commercial value of the other's identity." Tan, *Fame*, *supra* note 72 at 41.

[142] McGee & Scanlan, *supra* note 132 at 270.

[143] Reiter, *supra* note 96 at 708–709.

[144] Bass, *supra* note 16 at 807.

[145] *Midler*, *supra* note 29 at 463.

By contrast, both in theory and in practice, non-celebrities may invoke the personality right and personal data protection right to protect their interests in identity.[147] This is because neither of these rights is premised upon the protection of commercial value. Although the personality right encompasses the commercial exploration of identity,[148] its scope is considerably broader, safeguarding fundamental personal interest by ensuring the protection of human dignity and facilitating the free development of one's personality.[149] In doing so, it preserves the essential attributes of natural persons.[150]

Similarly, data protection laws – such as the GDPR – are deeply rooted in the principles of human dignity, autonomy, and freedom of will, and are primarily concerned with mitigating risks to the rights and freedoms of the data subject.[151] As neither personality rights nor personal data protection rights require the identity in question to possess commercial value, they are more readily accessible to non-celebrities than the right of publicity.

As AI technologies have rendered the voice of non-celebrities increasingly susceptible to cloning, personality rights and personal data protection rights offer more robust protection to non-celebrities than the right of publicity can afford. For instance, if a song employing an AI-generated voice clone of an unknown artist were to achieve commercial success and generate substantial profits, it would be exceedingly difficult for that individual to assert a claim under the right of publicity.[152] This is largely because the right of publicity is generally reserved for those whose identities possess recognised commercial value, a criterion seldom met by an unknown artiste. In such circumstances, the pursuit of a remedy under personality rights or personal data protection right – should such protections be available within the relevant jurisdiction – may offer a far more viable and effective course of action. These alternative legal frameworks may provide broader protection for individuals whose personal attributes have been exploited without their consent, irrespective of their prior recognition.

---

[146] J Thomas McCarthy, "The Human Persona as Commercial Property: The Right of Publicity" (1995) 19 Colum-VLA J L & Arts 129 at 133.

[147] McGee & Scanlan, *supra* note 132 at 266–269.

[148] Tilman Ulrich Amelung, "Damage Awards for Infringement of Privacy—The German Approach" (1999) 14 Tul Eur & Civ LF 15 [Amelung] at 36–37; Lauber-Rönsberg, *supra* note 96 at 813–816; Rösler, *supra* note 95 at 181; Reiter, *supra* note 96 at 693–694; Resta, *supra* note 95 at 45–46.

[149] Bass, *supra* note 16 at 828–829.

[150] Popovici, *supra* note 93 at 351–352.

[151] Claudia Quelle, "Enhancing Compliance under the General Data Protection Regulation: The Risky Upshot of the Accountability- and Risk-Based Approach" (2018) 9 EJRR 502 [Quelle] at 502.

[152] Shields, *supra* note 36 at 768–769.

### C. *Remedy*

All three rights offer certain remedies to complainants who are able to successfully substantiate their claims; however, these remedies differ considerably owing to the distinct natures of the respective rights. A successful claimant of the right of publicity may be granted injunctions and awarded statutory, compensatory, or, in some jurisdictions, even punitive damages.[153] As the right of publicity is fundamentally intended to protect the commercial value of an individual's identity, courts are typically required to assess the fair market value and the potential for future earnings associated with the complainant's attributes when determining the quantum of damages.[154] Furthermore, certain courts award damages for non-economic harm.[155] For instance, in the aforementioned *Waits* case, although the defendants argued that only monetary compensation for economic injury was permissible in actions concerning the right of publicity actions, the court disagreed and awarded Waits US$75,000 for injury to his personal interests, including humiliation, embarrassment, and mental distress.[156] Nevertheless, it should be noted that, in the absence of a federal statutory right of publicity, the scope and types of damages recognised by the courts vary considerably from one US state to another.[157]

By contrast, in most civil law jurisdictions, punitive damages are generally not available in cases concerning violations of personality rights. Nevertheless, in addition to injunctive relief, claimants may seek damages for both economic loss and non-economic harm – commonly known as non-material harm or moral damages – arising from factors such as pain and mental distress.[158] Notably, courts in a number of civil law jurisdictions, especially in Asia, may order the perpetrator to publish a corrective statement or a public apology, as mandated by law, to address the harm caused by an infringement of personality rights.[159] Such remedies are generally not available to claimants under the right of publicity.[160]

Much like the laws governing the right of publicity and personality rights, personal data protection regimes across jurisdictions have increasingly recognised damages for non-pecuniary loss. For instance, Article 82 of the GDPR explicitly provides for compensation for both material or non-material damages.[161] This principle was affirmed by the Court of Justice of the European Union in *UI v Österreichische Post AG* in 2023, where the court held that there is no threshold of seriousness required for non-material damages in data breach claims.[162] Likewise, the German Federal Court of Justice ("BGH") issued a decision in 2024 concerning the scraping of Facebook data. In this case, the BGH held that loss of control over personal data is sufficient to substantiate a claim for non-material damages,[163] determining that 100 euros was appropriate even in the absence of proof of financial loss.[164]

[153] Bergmann, *supra* note 95 at 498–500; Taylor Y Moore-Willis, "Locally Famous, Nationally Vindicated: A Comparative Analysis of Likeness-Based Recovery Avenues for Student-Athletes Under the Right of Publicity and Lanham Act" (2018) 24 BUJ Sci & Tech L 135 at 144–145.

[154] Moore-Willis, *supra* note 153 at 147–151.

[155] McGee and Scanlan, *supra* note 132 at 270–271.

[156] *Waits*, *supra* note 31 at 1103.

[157] Shields, *supra* note 36 at 769.

[158] Amelung, *supra* note 148 at 36–37; Bass, *supra* note 16 at 841–845; Bergmann, *supra* note 95 at 516–518; Daniel Biene, "Celebrity Culture, Individuality and Right of Publicity as a European Legal Issue" (2005) 36(5) IIC, 505 at 515; Jae Hyung Kim, "Protection of Personality Rights Under Korean Civil Law" (2017) 30 Colum J Asian L 131 [Kim] at 135; Kraski, *supra* note 95 at 949–952; Lauber-Rönsberg, *supra* note 96 at 809; Resta, *supra* note 95 at 50–51.

[159] Mauro Bussani and Marta Infantino, "Tort Law and Legal Cultures" (2015) 63 Am J Comp L 77 at 105–106; Robert H Hu, "Apology as an Intellectual Property Remedy in China: A Preliminary Examination of American Litigation Experiences" (2024) 51 Syracuse J Intl L & Com 177 at 180–182; Kim, *supra* note 158 at 135.

[160] Xuan-Thao Nguyen, "China's Apologetic Justice: Lessons for the United States?" (2013) 4 Columbia Journal of Race and Law 97 at 127–128.

Similarly, in *Liao v One Technical Culture Co. Ltd*, a case involving unauthorised processing of the plaintiff's personal data, the Beijing Internet Court awarded compensation for non-material damages,[165] invoking Article 69 of the PIPL. This provision stipulates that:

> [w]here the personal information processing infringes upon rights and interests relating to personal information and causes damage, and the personal information processor cannot prove that it or he is not at fault, the personal information processor shall assume liability for damage and other tort liability.[166]

Unlike the legal regimes governing the right of publicity and personality rights, personal data protections laws usually designate a government authority to enforce compliance. These authorities may initiate investigations following a complaint by a data subject.[167] The involvement of personal data protection authorities can offer significant advantages to data subjects, both in terms of gathering evidence and providing alternative avenues for enforcement. The GDPR, for example, empowers supervisory authorities to assist data subjects in exercising their right – such as the right to restriction of processing and right to erasure (*ie*, right to be forgotten) – where there has been unauthorised processing of personal data.[168] Additionally, each supervisory authority is vested the authority to issue administrative fines, the severity of which is determined by the circumstances of each case.[169] By way of illustration, in 2021, the Lithuanian Data Protection Authority imposed a fine of €20,000 on a sports club for violating the GDPR in its processing of biometric data.[170]

---

161 GDPR, *supra* note 112, Art 82.

162 *UI v Österreichische Post AG* [2023] (CJEU Case C-300/21).

163 Jakob Horn & Alexander Schmalenberger, "BGH: Loss of control over Facebook data justifies compensation – What companies must do now", *Taylor Wessing* <https://www.taylorwessing.com/en/insights-and-events/insights/2024/11/bgh-loss-of-control-over-faceook-data-justifies-compensation> (21 November 2024).

164 "Facebook users affected by data breach eligible for compensation, German court says", *Reuters* <https://www.reuters.com/technology/facebook-users-affected-by-data-breach-eligible-compensation-german-court-says-2024-11-18/> (18 November 2024).

165 *Liao v One Technical Culture Co Ltd* [2023] Beijing Internet Court Civ 3820 [*Liao*].

166 Personal Information Protection Law 2021, *supra* note 137, Art 69.

167 GDPR, *supra* note 112, Art 77.

168 *Ibid*, Arts 17 and 18.

169 *Ibid*, Art 83.

170 "Lithuanian DPA: Fine Imposed on Sports Club for Infringements of GDPR in Processing Fingerprints of Employees and Members", *European Data Protection Board* <https://www.edpb.europa.eu/news/national-news/2021/lithuanian-dpa-fine-imposed-sports-club-infringements-gdpr-processing_en> (2 June 2021).

In a similar vein, the PIPL in China empowers the relevant enforcement authorities with wide-ranging investigatory and rectification powers. These authorities are entitled to conduct on-site inspections, which may include the examination of equipment and articles implicated in personal information processing activities, and, where substantiated by evidence, the seizure or impoundment of equipment used for unlawful processing of personal information.[171] In addition, the regulatory body charged with personal information protection has the authority to order the responsible personal information processor to undertake corrective measures, issue formal warnings, and confiscate any illicit income derived from such contraventions. Where an application programme is found to be processing personal information in breach of the law, the enforcement authority may also mandate the suspension or termination of the provision of services by the offending application. Furthermore, the enforcement agency retains the discretion to impose administrative fines, the severity of which is determined in accordance with the gravity and particular circumstances of each individual case.[172]

Although all three types of rights offer mechanisms for victims to claim pecuniary and non-pecuniary damages, the remedies provided by personal data protection legislation appear to encompass a wider array of measures specifically designed to safeguard identity interests. The authority of data protection agencies to investigate infringements and impose substantial fines for non-compliance arguably affords the data subject a more robust layer of protection than that afforded by the right of publicity or the personality rights. Notably, the GDPR's right to be forgotten empowers individuals to request the erasure of their personal and sensitive data – including biometric identifiers – from the internet and relevant directories under certain circumstances.[173] Given that biometric identifiers are inherently distinctive, granting persons the capacity to have this data erased upon request effectively mitigates the risk of long-term misuse and supports the preservation of personal autonomy over biometric attributes.[174] Nevertheless, it must be observed that in certain respects, the right of publicity and personality rights offer even greater protection than personal data protection laws; these particular strengths will be explored in further detail in the subsequent two sections.

### D. *The Use of Complainant's Voice Data*

Although personal data protection laws confer certain advantages in safeguarding vocal identity – benefits not fully encompassed by regimes of right of publicity and personality right – these laws are not without their limitations. Such limitations, in fact, reveal the comparative strengths of protecting vocal identity under the right of publicity and personality rights frameworks. Broadly speaking, these shortcomings can be distilled into two primary areas: first, the requirement that the defendant must have utilised the complainant's actual voice, and secondly, the question of whether successors are entitled to assert claims on behalf of deceased voice owners against unauthorised users.

---

[171] Personal Information Protection Law 2021, *supra* note 137, Art 63.

[172] *Ibid*, Art 66.

[173] GDPR, *supra* note 112, Art 17.

[174] Llaneza, *supra* note 113 at 194–195.

Historically, prior to the advent of AI technology, prominent cases involving voice and the right of publicity often concerned a human imitator replicating the claimant's voice characteristics.[175] A notable example is the aforementioned *Midler* case, in which the US Court of Appeals for the Ninth Circuit determined that Ford Motor had infringed upon Bette Midler's right of publicity by engaging a singer whose voice bore a striking resemblance to Midler's own.[176] Although the defendants in these cases merely imitated the plaintiffs' voices, they did not make use of the plaintiffs' actual voice data.

Likewise, the personality rights doctrine does not require an infringer to use the plaintiff's actual recorded voice in order to establish infringement. For instance, in the German *Heinz Erhardt* case, the court found that the defendant's imitation sufficed to constitute a violation of personality rights.[177]

In contrast, where an individual seeks to invoke the personal data right over their voice, the legal framework requires that their specific voice data has been collected, processed, or used by another party. This stems from the underlying purpose of personal data regimes, which is to empower individuals to control their own personal data.[178] Consequently, the ambit of these rights is confined strictly to data relating to the subject themselves and does not extend to data about others. For instance, under China's PIPL, data subjects are vested with rights to be informed, to decide, and to restrict or refuse the processing of their "personal information."[179] Similarly, their entitlements to consult, duplicate,[180] transfer,[181] correct,[182] seek deletion,[183] and request explanation[184] are limited exclusively to their own personal information. Under the GDPR, the same principle holds true: the rights afforded to data subjects pertain solely to personal data concerning themselves. By way of illustration, Article 15 (the right of access) and Article 16 (the right to rectification) both explicitly delineate their application to data concerning the data subject only.[185] Therefore, absent direct access to a person's voice, the mere act of imitating – without any underlying processing of the actual data – does not, in itself, trigger issues of personal data protection.

In summary, when considering the protection of vocal identity, the right of publicity and the personality rights offer significantly broader safeguards than those afforded by personal data protection laws in this respect. Infringement under these regimes may be established even in circumstances where the claimant's actual data have not been employed, provided that identity confusion arises through imitation or the blending of other individuals' voice data. This scenario becomes particularly pertinent with the use of AI: for example, where a third party, whose voice happens to resemble that of the claimant, is engaged to provide voice samples for the training of an AI model that subsequently generates a voice indistinguishable from the original. This situation mirrors the recent controversy involving OpenAI and Scarlett Johansson, in which Johansson alleged the unauthorised use of her voice likeness. Regardless of jurisdiction or applicable law, if OpenAI's response – that no direct use of Johansson's voice data occurred – was accurate, it would nonetheless be considerably easier for Johansson to advance her claim under the right of publicity and the personality rights, as opposed to relying solely on the personal data protection right.

---

[175] Beard, *supra* note 16 at 1191–1192.

[176] *Midler*, *supra* note 29.

[177] *Heinz Erhardt*, *supra* note 33.

[178] Alexandra Giannopoulou Jef Ausloos, Sylvie Delacroix & Heleen Janssen, "Intermediating data rights exercises" (2022) 12 IDPL 316 at 319.

[179] Personal Information Protection Law 2021, *supra* note 137, Art 44.

[180] *Ibid*, Art 45.

[181] *Ibid*.

[182] *Ibid*, Art 46.

[183] *Ibid*, Art 47.

[184] *Ibid*, Art 48.

[185] GDPR, *supra* note 112, Arts 15 and 16.

Nevertheless, the personal data protection right's emphasis on actual use is not aways an obstacle to the protection of biometric information; rather, it serves as a double-edged sword. In situations where "recognisability" – the core element within both the right of publicity and the personality rights frameworks – is not established, the personal data protection right can function to protect the biometric data owner, thereby closing a loophole where neither the right of publicity nor personality rights would offer such protection. For example, if the voices of three individuals are mixed to produce a synthetic voice, it would be extremely difficult for any of them to make a successful claim based on the right of publicity or personality right if no single individual could be identified from the resulting synthetic voice. Consequently, they would not be able to argue that the synthetic voice falls within the protectable scope of their voices under either framework.

By contrast, because the personal data protection right focuses mitigating risks to the rights and freedoms of the data subject,[186] a breach of compliance requirements in itself can trigger the protection mechanism. For example, in 2025, the Beijing Internet Court issued a landmark case where the claimant's facial image was processed to create a face-swapping effect.[187] Even though the claimant's facial image data was indeed used, her identity could not be discerned from the resulting face-swapping images. The Court chose to apply data protection rights to protect the claimant's interest.[188]

In summary, personal data protection laws offer distinct advantages in safeguarding biometric identity, particularly through providing robust mechanisms to control the processing and use of one's own personal data. This is especially effective in circumstances where traditional concepts of recognisability are insufficient to trigger protection under the right of publicity or personality rights, thereby closing potential loopholes. In cases where no particular individual is recognisable from processed data, or where synthetic voices have been amalgamated beyond identification, the personal data protection right may provide a form of recourse where more traditional rights do not. Nevertheless, the scope of personal data protection legislation is inherently limited by its strict requirement that actual data concerning the data subject must have been used, collected, or processed. By contrast, the right of publicity and personality rights frameworks afford broader protection for vocal identity, encompassing acts of imitation and appropriation that do not involve the subject's actual recorded data, and permitting claims even where confusion arises solely by resemblance.

---

[186] Quelle, *supra* note 151.
[187] *Liao*, *supra* note 165.
[188] *Ibid*.

### E. *Posthumous Protection*

The use of AI technologies to clone the voices of the deceased has become increasingly widespread in recent years. New business models have emerged which offer AI-generated voice that closely resemble those of the deceased.[189] This trend of "digital resurrection" has given rise to significant legal and ethical concerns surrounding identity.[190] For instance, much to the shock and dismay of admirers, the director of the documentary *Roadrunner* used a cloned version of the late celebrity chef Anthony Bourdain's voice to simulate him reading aloud from his emails, deliberately choosing not to disclose the presence of AI in the film until after its release.[191]

While the deceased themselves cannot assert rights against unauthorised users, questions arise as to whether their successors may legally intervene. In this context, personal data protection rights appear more limited than the right of publicity and personality rights. The right of publicity is recognised as a transferable and inheritable property right,[192] a principle enshrined in statutes in certain states such as California, Illinois,[193] Alabama,[194] and Nevada.[195]

Most commentators have taken a position in favour of a postmortem right of publicity, whether as freely descendible right or subject to the condition of "lifetime exploitation".[196] Legislators have taken a similar stance. By 2025, a total of 27 states across the United States have recognised a postmortem right of publicity: 19 through statute, and eight by common law.[197] While a handful of states have yet to determine the issue, only one state – Wisconsin – has explicitly rejected any common law postmortem right of publicity, as held by the federal courts within this state. This position has not been overturned by state court decisions or by statute.[198]

---

[189] Jo Adetunji, "Talking to Dead People Through AI: The Business of 'Digital Resurrection' Might Not Be Helpful, Ethical… or Even Legal", *The Conversation* <https://theconversation.com/talking-to-dead-people-through-ai-the-business-of-digital-resurrection-might-not-be-helpful-ethical-or-even-legal-242404> (30 October 2024).

[190] *Ibid*.

[191] Helen Rosner, "The Ethics of a Deepfake Anthony Bourdain Voice", *New Yorker* <https://www.newyorker.com/culture/annals-of-gastronomy/the-ethics-of-a-deepfake-anthony-bourdain-voice> (17 July 2021).

[192] Jennifer E Rothman, "The Inalienable Right of Publicity" (2012) 101 Geo LJ 185 at 219.

[193] California Civil Code 1872, § 3344.1 (amended in 1999), Illinois Right of Publicity Act 1999, § 15.765, Illinois Compiled Statutes 1075/15.

[194] Alabama Code 1975 § 6-5-771(3).

[195] NRS § 597.790.

[196] McCarthy & Schechter, *supra* note 69 at § 9:9.

[197] *Ibid*, § 9:17.

[198] *Ibid*, § 9:18.

In *Factors Etc., Inc. v Pro Arts, Inc.*, where the name and image of the late rock musician Elvis Presley were misappropriated, the corporate licensee of Presley's publicity rights successfully obtained a preliminary injunction against the unauthorised use.[199] Although the US Court of Appeals for the Second Circuit refrained from establishing a general rule of inheritable publicity rights in this case, it held that the exclusive rights exercised during Presley's lifetime persisted after his death.[200] Likewise, in *Martin Luther King, Jr., Center for Social Change, Inc. v American Heritage Products, Inc.*, the Eleventh Circuit affirmed that "the right of publicity survives the death of its owner and is inheritable and devisable".[201] In *Crosby v HLC Properties, Ltd.*, the California Court of Appeal (Second Appellate District, Division Three) found that the original version of the relevant statute provided that the right of publicity was "freely transferable ... by contract or by means of trust or testamentary documents".[202]

Most civil law jurisdictions not only recognise the foundational existence of personality rights, but also uphold their persistence beyond death, embodying a deep-rooted tradition of respect for personal liberty, dignity, and reputation.[203] The posthumous protection of personality rights serves to preserve both the memory of the deceased and the honour of their heirs, spouse, and legatees.[204] Accordingly, under the doctrine of personality right, legal protection can be extended to the voice of an deceased individual, ensuring that their distinctive attributes remain safeguarded from unauthorised exploitation or misrepresentation.

For example, in the above *Heinz Erhardt* case, the son of the celebrated actor and author sought to prevent a radio advertisement in which an imitator not only impersonated Erhardt's voice but also used expressions and idiosyncrasies unmistakably associated with him.[205] The Higher Regional Court of Hamburg drew parallels with the protection of name and likeness, extending these protections to voice imitation.[206] In another German case concerning Marlene Dietrich, the court held that the commercial value of the general right to personality is descendible.[207] The court held that an individual's image, name, and other aspects of personality – particularly one's voice – may possess considerable economic value.[208] Accordingly, personality rights are regarded not only as protecting dignity but also as conferring pecuniary rights.[209] Thus, both posthumous economic interests, as exemplified by the *Marlene Dietrich* case, and moral interests, as illustrated by the *Heinz Erhardt* case, are safeguarded under the law of personality rights.[210]

---

[199] *Factors Etc., Inc. v. Pro Arts, Inc.*, 579 F 2d 215 (2nd Cir, 1978).

[200] *Ibid* at 222, Peter L Felcher & Edward L Rubin, "The Discernibility of the Right of Publicity: Is There Commercial Life after Death" (1980) 89:6 Yale LJ 1125 at 1125–1126.

[201] *Martin Luther King, Jr. Center for Social Change, Inc. v American Heritage Products, Inc.*, 694 F 2d 674 at 680–683 (11th Cir, 1983).

[202] *Crosby v HLC Properties, Ltd.* (2015) 223 Cal App 4th 597 at 608.

[203] Edwards & Harbinja, *supra* note 94 at 103–104.

[204] *Ibid* at 109, Eberle, *supra* note 95 at 1012.

[205] *Heinz Erhardt*, *supra* note 33.

[206] *Ibid*.

[207] Franz Hofmann, "The Economic Part of the Right to Personality as an Intellectual Property Right?" (2010) 2 *Zeitschrift für geistiges Eigentum* 1 at 15; Raymond Youngs, "Marlene Dietrich Case BGH 1 ZR 49/97", *The University of Texas at Austin Foreign Law Translations* <https://law.utexas.edu/transnational/foreign-law-translations/german/case.php?id=726> (1 December 1999).

[208] Franz Hofmann, "The Right to Publicity in German and English Law" (2010) 3 IPQ 325 at 330.

[209] *Ibid*.

Similarly, the Supreme People's Court in China established in 2000 that the image rights of natural persons are to be protected posthumously,[211] a principle further incorporated into Article 994 of the Civil Code.[212] In 2023, the Beijing Internet Court ruled on the unauthorised use of the image of a deceased soldier for advertising purposes.[213] The Court found that such use infringed upon the personality interests of the deceased and caused mental distress to the heirs, thereby awarding damages for both economic loss and mental distress and ordering a public apology.[214]

By contrast, approaches diverge in the sphere of personal data protection around the world regarding whether heirs may act in the interest of the deceased.[215] As the principal policy objective of personal data protection law is to safeguard individual's autonomy in shaping and conducting their own lives,[216] most personal data protection laws do not extend protection to the data of the deceased. The GDPR notably excludes the personal data of the deceased from its remit, although it permits member states to establish their own rules.[217] Countries, such as Belgium, Germany, Sweden, and the United Kingdom have similarly excluded the data of the deceased from protection under their laws.[218]

In contrast, approximately ten European jurisdictions – including Bulgaria and Switzerland – do provide for some protection of the personal data of deceased person, albeit limited in its scope and duration,[219] on the basis that such data remains linked to living individuals.[220] Likewise, China's PIPL offers similar protections when close relatives of the deceased can demonstrate that such protection serves their own lawful rights and interests.[221]

---

[210] Fennimore, *supra* note 131 at 405–407.

[211] Supreme People's Court of China, "最高人民法院关于周海婴诉绍兴越王珠宝金行侵犯鲁迅肖像权一案应否受理的答复意见" [Reply of the Supreme People's Court Concerning the Jurisdiction of and Whether to Accept the Case of *Zhou Haiying v Shaoxing Yuewang Zhubao Jinhang* (Jewellery Store) for Infringement of Lu Xun's Right to Image] *China Baike Court Publications Directory* <https://web.archive.org/web/20181122204903/https://www.chinabaike.com/law/zy/sf/fy/1337859.html> (26 June 2000, archived 22 November 2018).

[212] Civil Code 2020, *supra* note 101, Art 994: "[w]here the name, likeness, reputation, honour, privacy, or body of a deceased person is infringed upon, his spouse, children, and parents have the right to request the actor to assume the civil liability according to the law; and where the deceased has no spouse or children, and his parents are dead, other close relatives have the right to request the actor to assume the civil liability according to the law.".

[213] Huiyao Xu, "用已故老红军昔日采访照打广告，网店商家被判赔偿" [Online shop ordered to pay damages for using the photo of a deceased veteran Red Army soldier in advertising], *Beijing Daily* <https://news.bjd.com.cn/2023/06/08/10458037.shtml> (8 June 2023).

[214] *Ibid*.

[215] Egoyibo Lorrita Okoro, *Death and Personal Data in the Age of Social Media* (LLM Thesis, Tilburg University, 2021) <https://arno.uvt.nl/show.cgi?fid=147487> at 1.

[216] Mark-Oliver Mackenrodt, "Personal Data After the Death of the Data Subject—Exploring Possible Features of a Holistic Approach" in Mor Bakhoum, Beatriz Conde Gallego, Mark-Oliver Mackenrodt & Gintarė Surblytė-Namavičienė, eds, *Personal Data in Competition, Consumer Protection and Intellectual Property Law: Towards a Holistic Approach?* (Berlin: Springer, 2018) 273 at 279 [Mackenrodt].

[217] GDPR, *supra* note 112, Recital 27. This recital stipulates that: "This Regulation does not apply to the personal data of deceased persons. Member States may provide for rules regarding the processing of personal data of deceased persons."

[218] Edwards & Harbinja, *supra* note 94 at 114; Mackenrodt, *supra* note 216.

[219] Edwards & Harbinja, *supra* note 94 at 113–114; David Erdos, "Dead ringers? Legal persons and the deceased in European data protection law" (2021) 40 Computer Law and Security Review 105495 [Erdos] at 3.

In summary, heirs possess a more extensive suite of rights concerning the voice data of deceased persons under the regimes of right of publicity and personality rights than they do under most personal data protection frameworks globally. The significance of these rights is only likely to grow with the increasing popularity of AI-driven voice cloning technologies. At the same time, the limitations of personal data protection law – particularly in cases where the subject's actual voice data is not used, or where posthumous protection is rejected – stem from its intrinsically personal nature. Such law requires actual use of personal data by the infringer and limits protection primarily to the data subject themselves.

## VI. Conclusion

The human voice represents a distinctive form of biometric information, playing an indispensable role in social interaction and communication. Not only does the voice serve as a principal marker of personal identity, but it also underpins social relationships and facilitates the assimilation of information. Increasingly sophisticated AI-driven voice cloning technologies, however, threaten to erode this fundamental function. The remarkable ease with which AI models can now replicate individual voices has given rise to a host of profound social, legal, and ethical challenges.

Within contemporary legal systems, the right of publicity, personality rights, and the personal data protection right constitute the three principal legal frameworks for safeguarding the identity interests inherent in the human voice. Each of these rights is rooted in different legal traditions, possesses its own historical trajectory, and operates through distinct doctrines, requirements, and remedies.

The right of publicity is principally concerned with the commercial exploitation of identity and, in practice, its protection most commonly benefits celebrities and individuals whose personal attributes possess considerable economic value. This legal right enables such individuals to control and profit from the use of their voice, likeness, or other distinguishing features in commercial contexts.

In contrast, personality rights are founded upon the protection of the moral and dignitary interests inherent in personal identity. Unlike the right of publicity, personality rights are generally available to a broader spectrum of individuals, reflecting a more inclusive ethos. These rights are deeply rooted in the civil law tradition, which places a particular emphasis on safeguarding personal dignity and reputation. However, the protection afforded by personality rights remains largely limited to those jurisdictions with a civil law heritage and is not as widely recognised in common law systems.

Personal data protection law, meanwhile, generally offers more robust mechanisms for evidence collection and a broader range of protective measures than the other rights. Nevertheless, it proves less effective in certain scenarios – such as imitations where the subject's actual voice data is neither collected, processed, nor used, or in cases involving the unauthorised use of a deceased person's voice. These limitations stem from the fact that the personal data protection right maintains the strictest standard concerning the "personal" nature of the protected object, thereby narrowing its scope of application in comparison with the other two rights. However, this emphasis on actual personal data may also afford flexibility in situations where "identifiability" is not established in the synthetic voice.

[220] Erdos, *supra* note 219 at 1.

[221] Personal Information Protection Law 2021, *supra* note 137, Art 49.

Although these three legal rights share the common objective of safeguarding the interests that underpin vocal identity, they diverge significantly in their scope, effectiveness, and practical applicability. As AI-driven voice cloning technologies continue to advance, the unique strengths and inherent limitations of each right become increasingly apparent and pronounced. A careful comparative analysis of these legal frameworks not only illuminates the relative efficacy of each approach in addressing the novel challenges posed by AI but may also provide a valuable foundation for future reforms.